# A Deep Learning Framework for Hydrogen-fueled Turbulent Combustion Simulation

Jian An[a,b], Hanyi Wang[c], Bing Liu[a], Kai Hong Luo [b*], Fei Qin[a], Guo Qiang He[a],

[a] Science and Technology on Combustion, Internal Flow and Thermal-structure Laboratory, Northwestern Polytechnical University, Xi'an Shaanxi 710072, PR China

[b] Department of Mechanical Engineering, University College London, London WC1E 7JE, United Kingdom

[c] Center for Combustion Energy, Tsinghua University, Beijing 100084, China

**Corresponding Authors[*]:**   Kai Hong Luo      E-mail: k.luo@ucl.ac.uk

**Abstract**

The high cost of high-resolution computational fluid/flame dynamics (CFD) has hindered its application in combustion related design, research and optimization. In this study, we propose a new framework for turbulent combustion simulation based on the deep learning approach. An optimized deep convolutional neural network (CNN) inspired from a U-Net architecture and inception module is designed for constructing the framework of the deep learning solver, named CFDNN. CFDNN is then trained on the simulation results of hydrogen combustion in a cavity with different inlet velocities. After training, CFDNN can not only accurately predict the flow and combustion fields within the range of the training set, but also shows an extrapolation ability for prediction outside the training set. The results from CFDNN solver show excellent consistency with the conventional CFD results in terms of both predicted spatial distributions and temporal dynamics. Meanwhile, two orders of magnitude of acceleration is achieved by using CFDNN solver compared to the conventional CFD solver. The successful development of such a deep learning-based solver opens up new possibilities of low-cost, high-accuracy simulations, fast prototyping, design optimization and real-time control of combustion systems such as gas turbines and scramjets.

# 1. Introduction

With the rapid development of high-performance computers, computational fluid/flame dynamics (CFD) can produce high-resolution simulation results that are of increasing importance to academic research and industrial R & D. However, the process of solving the partial differential equations for turbulent combustion with realistic chemistry is computationally expensive. Computing costs are still the biggest obstacle, when engineers have to repeatedly do simulations to evaluate the performance of a combustor or optimize a design for different working conditions. This is especially true in the design of complex combustion systems such as gas turbines, ramjets, scramjets and rocket engines. One of the alternatives is to construct surrogate models to act as a quick and specialized solver. Over the past ten years, deep learning methods have become popular and are regarded as an effective alternative to offer the possibility for researchers to make a trade-off between a model's precision and computational time.

The capability of deep learning has attracted many researchers' attention for the past few decades. Its development resulted from the great progress in the artificial neural network (ANN), which can effectively deal with nonlinear relationships. Benefiting from the improvement of algorithms [1] and the emergence of various types of ANNs designed for different purposes [2], ANN's applications in the field of deep learning have gradually expanded to the industry, e.g. emission control of engines [3, 4] and modeling of engine characteristics [5, 6].

Many attempts have been made to apply ANNs to fluid dynamics simulation [7]. To model the turbulence, Tracey et al. [8] presented a landmark proof-of-concept of a new ANN approach to build a representation of turbulence modeling closure terms using supervised learning algorithms. Ling et al. [9] employed a deep neural network technique to predict the Reynolds stress anisotropy tensor. Wang et al. [10] and Wu et al. [11] proposed a comprehensive perturbation strategy to train and revise

Reynolds stresses from a DNS database. These results are both exciting and encouraging and the applications of deep learning in turbulence modeling are systematically reviewed by Duraisamy et al. [12]. In addition to turbulence, application of deep learning in combustion is the focus of this study. The developed methods for simulation of turbulent combustion mainly include one or more of the following three kinds of strategies. 1) Chemistry representation. 2) Subgrid scale modeling / combustion modeling. 3) Surrogate/specialized solver.

One of the researchers' original intentions of applying machine learning to combustion is to deeply explore results with reduced computational resources. As the calculation of reactions consumes the majority of the computational resources, most efforts are devoted to representing chemical reactions by using different topologies of ANN. The application of machine learning in combustion starts from reproducing species changes in a three-step chemistry mechanism for hydrogen turbulent combustion [13]. Subsequently, Blasco et al. [14] applied the ANNs to a reduced, 5 steps and 8 species mechanism for methane–air combustion. They proposed a two-layer network to divide the samples of the thermodynamics space into sub-domains by using Self Organizing Map (SOM) and use multiple ANNs with the aim to get a dedicated ANN for each sub-domain. The SOM-ANN topology was tested by simulating a Partial Stirred Reactor (PaSR) and encouraging results were reported. Furthermore, Chatzopoulos and Rigopoulos [15] and Franke et al. [16] developed a method of combining the Rate-Controlled Constrained Equilibrium (RCCE) with the SOM-ANN for a mechanism of CH4–air combustion with 16 species. After training, the simulation results show that RCCE-SOM-ANN topology the ANN approaches have the ability to reduce the computational cost by one to two orders of magnitude. Moreover, some efforts have been made to use artificial neural networks to accelerate the global sensitivity analysis of chemical mechanisms and also exhibits better performance in convergence and stability comparing with conventional methods [17, 18].

ANNs have also been demonstrated for their power in reformulating and predicting subgrid-scale models for combustion modeling, which is the main challenge in turbulent combustion modelling [19]. For example, the flamelet models based on tabulations of small flamelet has recently become popular for its efficiency [20]. Kempf et al. [21] applied feed-forward ANNs to represent a steady flamelet model in LES with the outputs of species mass fractions, density and viscosity as a function of the mixture fraction. A multi-layer perceptrons with two hidden layers were employed in the research. Compared with conventional flamelet table integration approaches, the ANNs-based tabulation can reduce memory by approximately three orders of magnitude with unchanged CPU cost. Similarly, Emami et al. in [22] conducted an analogous study by using an optimal ANN-based flamelet framework to model the turbulence-combustion interaction in a turbulent CH4/H2/N2 jet diffusion flame for RANS. Compared with numerical integration for the estimation of mean thermo-chemical variables in their computational fluid dynamics code, the ANN method yields good predictions with decreased computational cost.

In addition, some efforts have been made in a different context to utilize ANN for combustion modelling. Sen et al. in [23] evaluated the applicability of ANNs approach as LES sub-grid model for the calculation of chemical reaction rates in reactive flows. With a well-trained architecture, the ANN can successfully predict the reaction rates with an accelerated calculation speed and reduced memory. In a subsequent study, Sen et al. [24] trained the ANNs with data generated from stand-alone linear eddy mixing (LEM) simulations instead of those from DNS of a laminar FVI results as in [23]. Results suggest that the ANNs trained by LEM results are able to grasp the correct flame physics with same accuracy as direct integration could offer, but with reduced memory and accelerated calculation speed. The integrated LEM-ANN LES framework was further applied to a flame-turbulence interaction near extinction and reignition of a non-premixed, syngas/air flame, contributing to about 5-fold speedup as that from a direct integration [24].

Furthermore, Nikolaou et al. [25] developed a data-driven modeling framework for estimating progress variables in LES. In their research, a DNS-data-trained CNN was employed to perform deconvolution on two variables, i.e. the filtered density and the filtered density-progress variable product. The two variables were further used to approximate the sub-grid scale progress variable variance and the filtered reaction rate. Inspired by a U-net technique (a type of CNN for image segmentation), Lapeyre et. al [26] applied a CNN method to simulate the subgrid scale flame surface density for LES. The training data set is built by DNS data of a premixed turbulent flame stabilized in a slot-burner configuration. After training, the CNN-based model was then applied to estimate subgrid scale wrinkling and validated by an unsteady turbulent flame. Very recently, Seltz et al. [27] proposed a unified modelling framework for all unresolved terms in the filtered progress variable transport equation in LES of turbulent premixed flames. The CNN was deployed and trained by a DNS database of a turbulent premixed stoichiometric methane/air jet flame. The results showed that the trained networks had been shown to produce quantitatively good predictions of all unresolved terms in an a priori study without having to resort to solving any additional transport equations. These studies have witnessed the excellent predictive performance of ANNs/CNNs in turbulent combustion sub-grid modeling.

Because of the above successful applications, an emerging area of research is to build specialized solvers, which is also the point of interest in this paper, but there are only limited results. Rajabi et al. [28] pioneered the use of a multivariate Levenberg-Marquardt neural network (LMNN) to predict the turbulent flow over a backward-facing step. The LMNN was informed by using DNS data and can then be used as a fast surrogate solver to accurately estimate the velocity field. Guo et al. [29] proposed a general and flexible approximation model for real-time prediction of non-uniform steady laminar flow in a 2D or 3D domain based on CNNs and successfully reduced the total consumed time to two orders less than a GPU-accelerated CFD solver. Zhang et al. [30] probed an approach to predict the airfoil lift coefficient by training a CNN model of multiple flows with various Mach numbers and Reynold numbers. The model exhibits a competitive prediction accuracy in predicting lift coefficients for unseen airfoil shapes. In summary, unlike the first two categories, this paradigm considers the flow field as a whole picture, resulting in a more efficient specialized solver for a problem. This kind of solver is crucial in R & D such as rapid prototyping, optimal design. However, the above studies only focus on turbulent flow and there is no such a precedent for specialized solvers of turbulent combustion as far as we know.

In this paper, we explore the feasibility of deep learning to predict turbulent combustion and construct a specialized solver in terms of physical properties and species distribution. An optimized architecture was developed and then validated in a hydrogen flame in a cavity in the terms of accuracy and efficiency.

## 2. Methodology

### 2.1 Overview

The results of numerical simulation of combustion are usually obtained by iteratively solving a large number of equations, whereas the deep learning technique provides an alternative way to reach the same end goal at a significantly reduced complexity and computational cost. Deep ANN has been shown to approximate any complex continuous functions [31], without actually solving the set of equations governing combustion. In this study, a framework of a specialized solver for combustion simulation based on an optimized deep convolutional neural network is proposed. Furthermore, the approach is applied to establish a fast solver for turbulent combustion in a cavity. Since deep learning is a purely data-driven approach that deconstructs the data in its own way, which is not based on any prior-knowledge, data generation is needed for the neural network. In the following, we describe the key components in turn.

**2.2 Deep learning architectures**

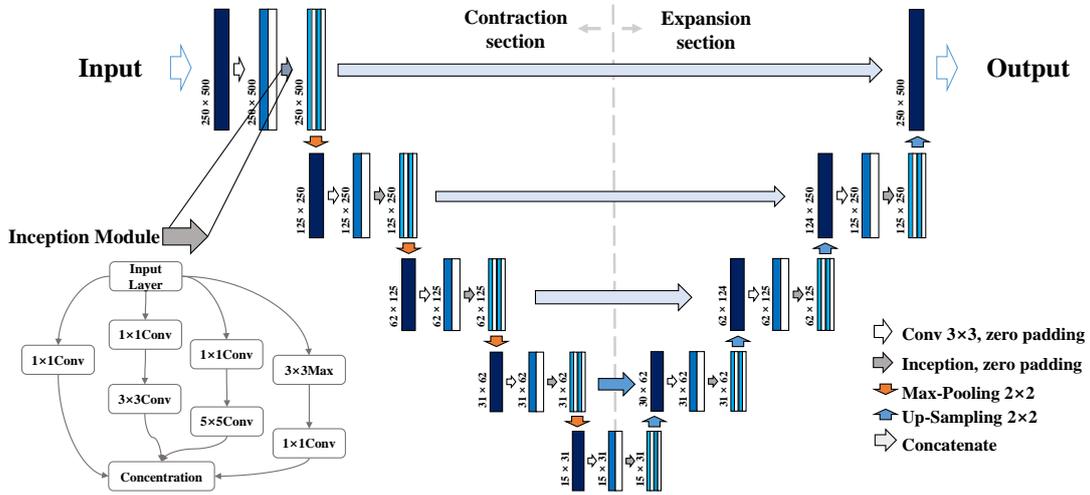

**Fig. 1.** Architecture of CFDNN framework.

Inspired by the U-net approach, we aim to develop a deep learning architecture for combustion, named CFDNN. U-net was first proposed by Ronneberger et al. [32], and has been proven to be an effective tool for complex segmentation tasks in the area of biomedical imaging through accurately extracting the characteristics of input data at different scales [33, 34]. Furthermore, U-net can not only convert the feature map into a vector, but also reconstruct the feature from this vector while greatly reducing sample requirements. Therefore, when considering the complex multiscale problem of combustion, a natural choice would be to us U-net as the basis and some successful attempts have been made [26, 35]. However, the original U-net is meant for a segmentation task, the output layer is designed to represent a categorical distribution. In the present combustion case, the U-net was adapted for the regression task by rectifying linear unit activation in the output layer. Furthermore, this structure has been further expanded upon by embedding the inception modules [36]. The concept of the Inception was firstly proposed by a group from Google, where they discussed how to extract more features with the same amount of computation. As shown in Fig. 1 (lower left), it is composed of parallel paths with different sizes convolutional layers. When input is coming in, these multiple paths can extract features of different scales than a single path without increasing too much computation.

Finally, as shown in Fig. 1, the newly developed network consists of the contraction and expansion sections. The contraction section can effectively learn complex features by step-by-step dimensionality reduction and decomposition using the inception module. It is composed of parallel paths with convolutional layers of different sizes. When input is coming in, these multiple paths can extract features of different scales rather than a single path, without increasing much computational load. After that, all feature maps at different paths are concatenated together as the input of the next module. This section can therefore be formulated as:

$$ConvNN : D_i = \{\varphi_i(x, y, z)\} \rightarrow H_i^k = \{f(x, y, z)\} \tag{1}$$

where $(x, y, z) \in \mathbb{R}^3$ is the 3-dimensional coordinate system of the simulation. $i$ is an index for the flow field variables, e.g. temperature, velocity, mass fraction of species etc. $ConvNN$ denotes the mapping from the simulation result to a convolutional space with $k$ features (user defined parameter) using multiple convolution.

The heart of this architecture lies in the expansion section, which, as you can see, is completely symmetric with the contraction section, and can be represented as:

$$DeConvNN : H_i^k \rightarrow \tilde{D}_i = \{\tilde{\varphi}_i(x, y, z)\} \tag{2}$$

The prediction of field variables $\tilde{\varphi}_i$ is extracted by the $DeConvNN$ operation. This action would ensure that the features that are learned while contracting the image will be used to reconstruct it. After these local optimizations, the overall framework requires a much smaller number of samples with acceptable amount of computation, allowing us to complete training and simulation even on a personal computer.

$$MSE = \frac{1}{n}\sum_{n=1}^{n}\left(D-\tilde{D}\right)^2 \qquad (3)$$

The training procedure is an optimization for minimizing the mean squared error (MSE) between $\tilde{D}$ and $D$, the Eq. (3). In this study, the Adam optimizer [37] is used to do this optimization. Adam is an extension of stochastic gradient descent (SGD) algorithm. The method is computationally efficient and straightforward to implement and has been proved to have good performance in processing large amounts of data and parameters [38, 39]. The training of the network was conducted using the TensorFlow (www.tensorflow.org) library.

## 3. Application of CFDNN

### 3.1 Data preparation

In this study, turbulent combustion in a cavity was chosen as an example for demonstrating the capability of the new proposed framework. The cavity commonly serves as a flame stabilizer in combustors, especially for transonic and supersonic combustion systems [40, 41]. It is also a simplified prototype of gas turbine combustors. This configuration contains complex physical phenomena, such as strong interaction between turbulence and combustion, ignition, etc. Therefore, unsteady RANS simulations are carried out to generate data sets for CFDNN. Figure 2 shows the detailed configuration of the chosen cavity. At the inlet, a hydrogen jet is injected into the combustion chamber. For hydrogen combustion, a widely used 9-species mechanism [42] was employed.

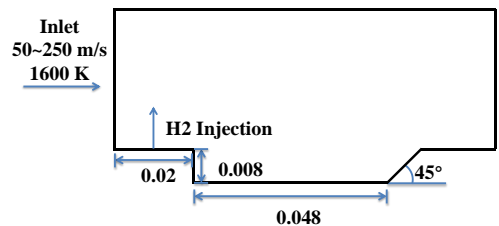

**Fig. 2.** Configuration of the cavity.

There are many parameters that can be changed as variables. Currently, as a proof of concept, the data set for training was generated by only changing the inlet velocity. The inlet velocity is varied in the range of 50 ~ 202 m/s with a step change of 4 m/s (i.e. …54, 58, 62…). Finally, the flow field data (including temperature, pressure and mass fraction of 9 species) of 38 cases are converted into matrixes, forming the training set. Furthermore, a data set for testing the performance of CFDNN was also generated with an inlet velocity of 45 to 210 m/s and a step change of 5 m/s (i.e. …50,55,60…), resulting in a total of 25 cases (excluding the same study as the training set).

All simulations were conducted with the OpenFoam platform. The partially stirred reactor (PaSR) combustion model and the k-Omega SST model were employed to model turbulence-chemistry interaction and turbulence, respectively. The time step is set to $2\times10^{-8}$ s, resulting a maximum CFL number < 0.3.

**3.2 Training and simulation**

Figure 3 illustrates how the information of combustion flow field is fed into CFDNN for training. In order to achieve the goal of predicting combustion, all flow field data, including temperature, velocity and mass fraction distribution of the 9 species, are converted into a vector matrix, and then sample pairs (consisting of inputs and targets for inputs) are formed in a specific order. For example, the input at $t_0$ is the flow field data at time $t_0$, while the target is the flow field information at the next time step $t_0 + \varDelta t$, which is then transmitted back to CFDNN as another input.

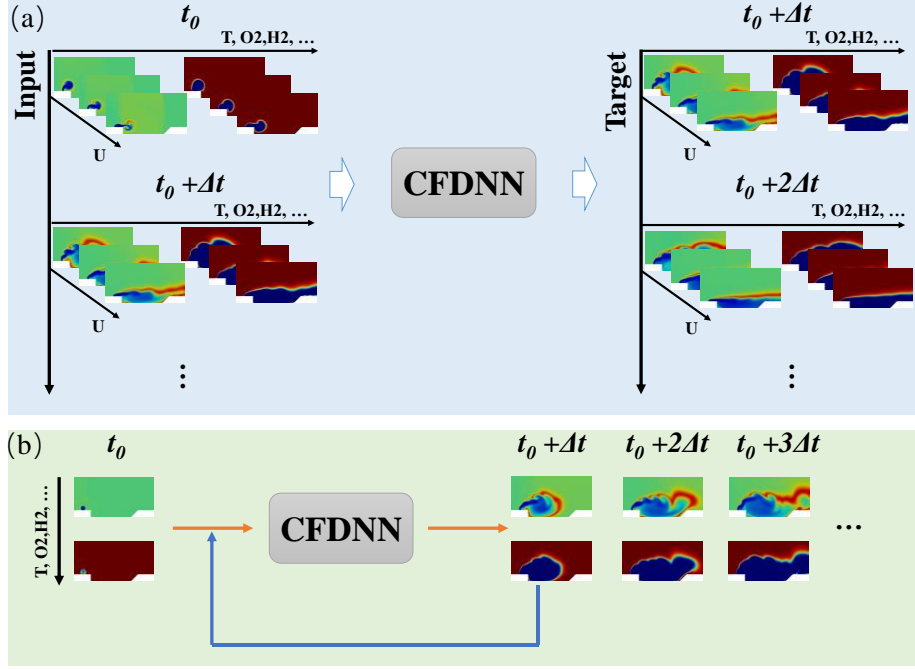

**Fig. 3.** Illustration of (a) training and (b) simulation procedures of CFDNN.

Moreover, the data set is divided into two parts, a training set (80%) and a validation set (20%). The validation set is used to check if an overfitting or underfitting has occurred. This is the so-called cross-validation technique commonly used in deep learning. When the MSE is less than $1\times10^{-4}$, the training is considered complete and a CFDNN-based solver for this situation is established. Finally, we tested the solver with the 25 unseen cases, the results of which are discussed in the next chapter. It is worth noting that, as shown in Fig. 3 (b), the simulation process, unlike the training process, only needs to input data of $t_0$ to complete the prediction of the entire time series. Because the prediction obtained in the previous step can be used as the input in the next step.

## 4. Results and discussion

The main motivation for our CFDNN-based solver is that CNN prediction of unsteady combustion is considerably faster than traditional solvers, potentially allowing the development of a rapid design tool and real-time simulation. In this section, the performance of the new method, including accuracy and efficiency, is fully demonstrated.

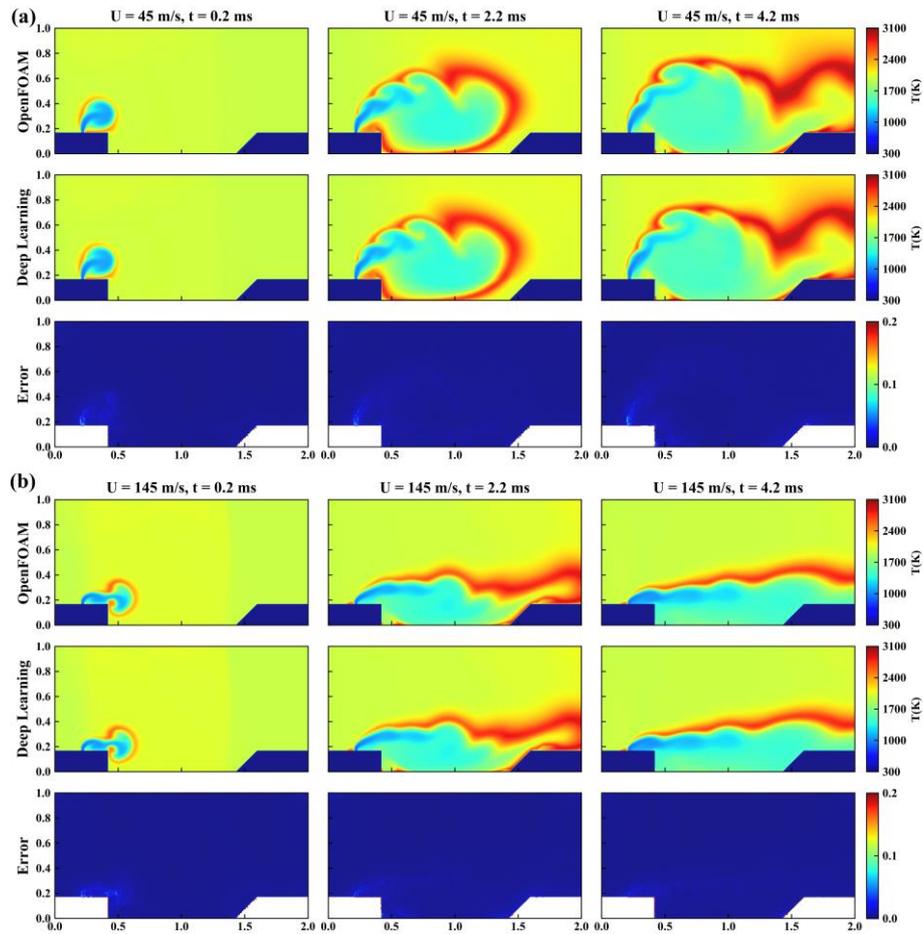

**Fig. 4.** Comparison of temperature distributions between OpenFoam and CFDNN results at inlet velocity (a) 45 m/s; (b) 145 m/s.

Figure 4(a) shows the temperature distribution of several moments obtained by Openfoam and CFDNN when the inlet velocity is 45 m/s. It is important to note that this inlet velocity is not within the range included in the training set. The results show that CFDNN is comparable to OpenFoam in that it accurately calculates the temperature distribution and captures the curvature of the flame surface caused by turbulence throughout the flow field. Meanwhile, from the perspective of time series, CFDNN well captures the process of flame ignition, development and finally formation of a stable turbulent flame under the influence of turbulence. In this case, since the inflow velocity is low, the fuel jet penetration depth (The vertical axis value of the hydrogen jet when the horizontal axis = 0.5) is large, and thus a strong interaction with the mainstream occurs. However, when the inlet velocity rises to 145 m/s, as shown in Fig. 4(b), the penetration depth is greatly reduced, and the flame needs to be stabilized by the cavity to continue to develop. High consistency between the OpenFoam and CFDNN results is shown. The third row of the two figures shows the difference between the two by introducing a percentage error, which is defined as follows:

$$Error = \frac{|\varphi_{OpenFoam} - \varphi_{CFDNN}|}{\varphi_{OpenFoam}} \quad (4)$$

The maximum difference between the OpenFoam and CFDNN results is less than 0.2%. The error distribution illustrates that the difference between the two is mainly in the small area of the flame root. This is mainly due to the large gradient in a relatively small area, which is tricky for feature recognition.

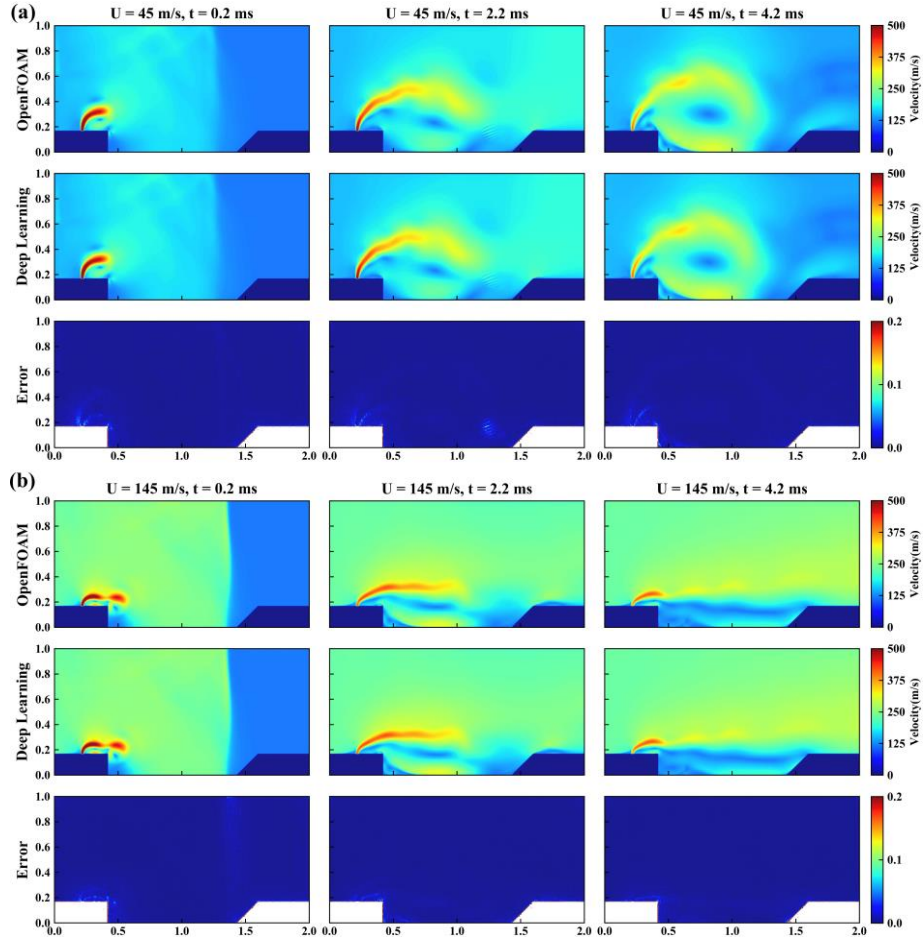

**Fig. 5.** Comparison of velocity distributions between OpenFoam and CFDNN results at inlet velocity

(a) 45 m/s;(b) 145 m/s.

Figure 5 plotted the velocity distributions at the two inlet velocities. The position of the airflow front during the unsteady process of high-speed airflow entering the cavity was accurately simulated by CFDNN as well as OpenFoam, which indicates the possibility of predicting the shockwave.

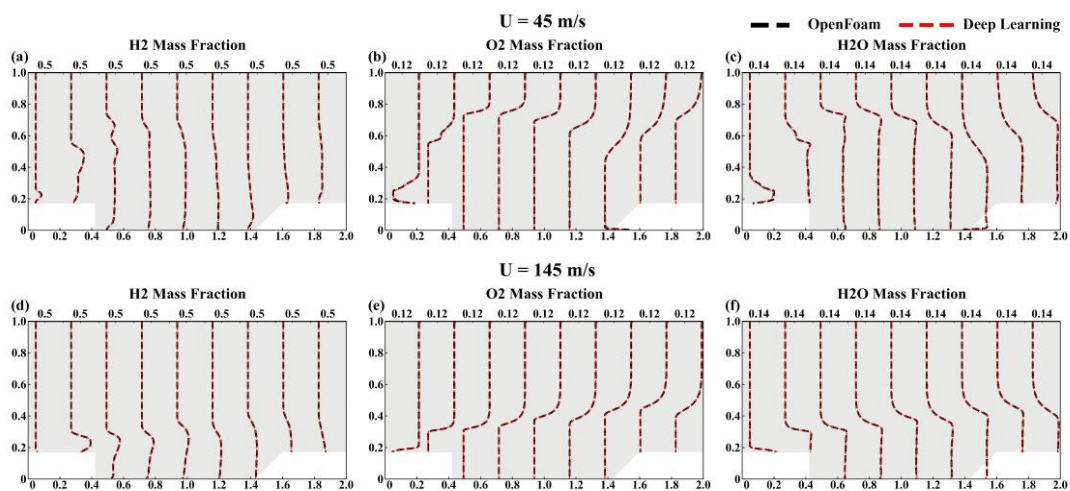

**Fig. 6.** Comparison of OpenFoam and the CFDNN results of mass fraction of selected species at inlet velocity (a)~(c) 45 m/s; (d) ~ (f) 145 m/s.

A key point of this paper is to realize the prediction of the mass fraction distribution of species. The mass fraction profiles of three representative species ($H_2$, $O_2$ and $H_2O$) at the two inlet velocities were selected and shown in Fig. 6. The time instant is selected at t = 4.2 ms because the flow has entered a steady state. Due to page limitations, the detailed contours are displayed in the supplementary material. Again, the result of CFDNN is highly consistent with that of OpenFoam, demonstrating the excellent performance of CFDNN in predicting the distribution of mass fraction. Furthermore, as the inlet velocity increases, the consumption of $O_2$ and the generation of $H_2O$ gradually shifted to the inside of the cavity. This indicates that as the combustion is increasingly concentrated within the cavity, the role of the cavity as a flame holder becomes more and more indispensable. Besides, the contours of OpenFoam and the CFDNN results of mass fraction of selected species at inlet velocity 45 m/s and 145 m/s in the Supplementary material. The results at 210 m/s were also given in the file, which supports the above speculation.

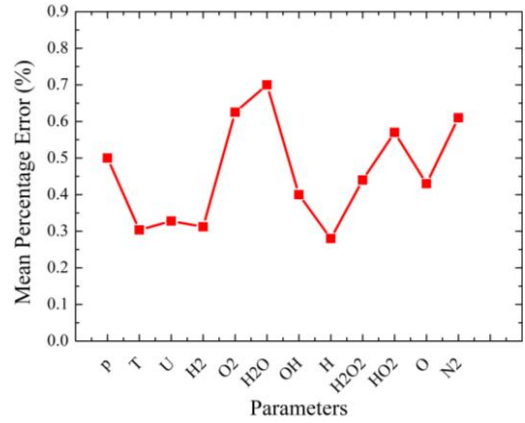

**Fig. 7.** Average percent error of all parameters.

Moreover, the average percent error of each parameters in all testing cases is displayed in Fig. 7 to evaluate the accuracy of CFDNN. The error of all parameters is less than 1%, which is completely acceptable.

**Table 1** Computational cost of two methods.

|  | OpenFoam | CFDNN |
|---|---|---|
| Training (s) | - | 14400 (24 cores) |
| Simulation (s) | 5799 (24 cores) | 51 (4 cores) |

In addition to accuracy, computational cost is another concern. Table 1 listed the costs for the two methods at the training and simulation phase. The new CFDNN framework requires advanced training, which is computationally intensive. However, once the training is finished and the deep learning solver is established, CFDNN can simulate a complete time series of combustion at hundreds of times faster than traditional methods. This feature will provide tremendous support for scenarios that require real-time calculation, such as rapid design, and even real-time control of combustion systems.

## 5. Conclusions

In this study, a framework of a deep learning-based solver for combustion simulation was established, which was realized by integrating an optimized deep convolutional neural network, CFDNN. A hydrogen flame in a cavity was used as an example to demonstrate the performance of the proposed method in terms of both accuracy and efficiency. By training CFDNN with data from the cavity flame at several different inlet velocities, CFDNN demonstrated the feasibility of simulating the results both within and even beyond the range of the inlet velocity of training data. The CFDNN results of both spatial distributions and temporal dynamics show excellent agreement with OpenFoam results. In terms of efficiency, CFDNN is characterized by bringing the huge amount of computation forward to the training phase, while reducing the cost of simulations by hundreds of times.

To the best of the author's knowledge, this is the first application of deep learning to reconstruct turbulent combustion field. Generally speaking, our study can be seen as a proof of concept, demonstrating that the prediction of turbulent combustion and species distributions can be reformulated as a machine learning problem. The application of this framework will make real-time numerical simulation possible, thus accelerating the development of technologies such as digital twins, rapid prototyping and emission control, with applications to complex systems such as gas turbines and scramjets. The robustness and versatility of the proposed CFDNN framework will be further tested for a wide range of combustion problems in follow-on research.

**Acknowledgments**

Support from the UK Engineering and Physical Sciences Research Council under the project "UK Consortium on Mesoscale Engineering Sciences (UKCOMES)" (Grant No. EP/R029598/1) and the China Scholarship Council (CSC) is gratefully acknowledged.